\documentclass[12pt]{article}
\usepackage{amsmath,amsfonts,amssymb}
\usepackage{hyperref}
\usepackage{graphicx}
\usepackage{slashed}
\usepackage{color} 
\usepackage{epsfig}
\usepackage{psfrag}
\makeatletter\@addtoreset{equation}{section}\makeatother

\begin{document}
\bibliographystyle{utphys}

\hfill {\tt UT-KOMABA-17-4}\\
\vspace{-4mm}

\begin{center}
{\LARGE Comments on supersymmetric renormalization }\\
\vskip 4mm
{\LARGE  in two-dimensional curved spacetime}
\end{center}

\vspace{5mm}

\begin{center}

\renewcommand{\thefootnote}{$\alph{footnote}$}

Takuya Okuda

\vspace{3mm}
University of Tokyo, Komaba \\
Meguro-ku, Tokyo 153-8902, Japan \\
\vskip 1mm
\href{mailto:takuya@hep1.c.u-tokyo.ac.jp}{\tt takuya@hep1.c.u-tokyo.ac.jp}

\renewcommand{\thefootnote}{\arabic{footnote}}
\setcounter{footnote}{0}

\end{center}

\vskip5mm

\abstract{
In this technical note we introduce a manifestly gauge-invariant and supersymmetric procedure to regularize and renormalize one-loop divergences of chiral multiplets in two-dimensional $\mathcal{N}=(2,2)$ theories in curved spacetime.
We apply the method, a version of Pauli-Villars regularization, to known examples of supersymmetric localization and demonstrate that the partition functions are renormalized.
}

\tableofcontents

\section{Introduction: supersymmetric regularization and countererms for a chiral multiplet}
\label{sec:supersymm-regul-supe}

In this note we explain how a version of Pauli-Villars (PV) regularization can be used for the one-loop determinant of an $\mathcal{N}=(2,2)$ chiral multiplet coupled to background gauge and gravity multiplets.
The relevant facts about $\mathcal{N}=(2,2)$ rigid supersymmetry and supergravity are summarized in Appendices \ref{sec:review-mathcal-n=2} and \ref{sec:full-supergr-with}, respectively.

Though the gauge and the gravity multiplets are treated as non-dynamical here, one may promote them to dynamical fields.
An application is discussed in  \cite{HLO}.

\subsection{Pauli-Villars regularization}
\label{sec:supersymm-pauli-vill}

A simple choice of supersymmetric regularization is a version of Pauli-Villars regularization.%
\footnote{%
A similar regularization method would be to add fictitious chiral multiplets with the usual statistics and let them couple to $U(1)_\text{F}$ in such a way that the charges add up to zero.
In order to make the regulator multiplets heavy, we introduce another symmetry $U(1)_\text{extra}$, under which the original fields are neutral but the charges of the regulator fields are non-zero and sum to zero.
We would then use $U(1)_\text{extra}$ to introduce a large twisted mass and integrate out the regulator multiplets.
We should be able to remove the divergences by supergravity counterterms.  
See \cite{Doroud:2012xw} for a use of similar regularization.
A difference from the PV method is that bosonic and fermionic determinants are not separately regularized.
}
Let us consider a chiral multiplet with charge $+1$ for the flavor symmetry group $U(1)_\text{F}$ and the complex central charge ${\rm z}$.
Often in a supersymmetric background, the complex scalar $\sigma$ in the non-dynamical $U(1)_\text{F}$ vector multiplet is constant; in that case $\sigma$ is called the twisted mass.
In order to regularize the one-loop amplitudes, we add to the action supersymmetric kinetic terms for extra $2N_\text{PV}-1$ chiral multiplets such that $ N_\text{PV} - 1 $ of them have the usual statistics, and the remaining $N_\text{PV}$ have the opposite statistics.
We let the index $J$ run from $0$ to $ 2 N_\text{PV} - 1 $, and $j$ from $1$ to $ 2 N_\text{PV} - 1 $.
We use $\epsilon_{J}=\pm 1$ to denote the statistics, $+1$ and $-1$ for the usual and the opposite, respectively.
We set $\epsilon_0=+1$. 

\begin{center}
  \begin{tabular}{l|l|lll}
    &$\epsilon_J=+1$ & $\epsilon_J= -1$\\
    \hline
    scalar component & boson & fermion \\
    spinor component & fermion & boson \\
    auxiliary field component & boson & fermion \\
  \end{tabular}
\end{center}

Let the $j$-th multiplet have an integer charge $b_j$ under  $U(1)_\text{F}$.
In order to make the regulator fields very heavy, we couple them to another vector multiplet for an abelian symmetry $\widetilde{U(1)}_\text{PV}$, under which the $j$-th PV multiplet has charge $a_j$.
We take this symmetry group to be non-compact, $\widetilde{U(1)}_\text{PV}=\mathbb{R}$ rather than $U(1)$, so that $a_j$ do not have to be an integer.%
\footnote{%
Alternatively, we can introduce for each $j$ a $U(1)$ vector multiplet whose scalar component is identified with $a_j\Lambda$.
This leads to the identical expressions for regularized partition functions.
}
For convenience we take $a_j$ to be positive: $a_j>0$.
For the vector R-symmetry, we assign R-charge $c_j q$ to the $j$-th PV multiplet.
We set $a_0=0$, $b_0=1$, and $q_0=q$.
We denote the complex scalar in the  $U(1)_\text{PV}$ vector multiplet by $\Lambda$.
As we will see, $\sigma$ and $\Lambda$ can be constants or non-trivial functions in spacetime, depending on the supergravity background.

\begin{center}
  \begin{tabular}{c|cc}
    $J=0$&$J=j\in\{1,\ldots,2N_\text{PV}-1\}$ \\
    \hline
    physical & unphysical (PV ghosts) \\
    $\epsilon_0=+1$ & $\epsilon_j = \pm 1$ \\
    $a_0 = 0 $ & $ a_j \in \mathbb{ R} -\{0\}$ \\
    $b_0= +1$ & $b_j \in \mathbb{Z}$ \\
twisted mass $\sigma$ & twisted mass $a_j \Lambda + b_j \sigma$ \\
R-charge $ q_0 = q$ & R-charge $ c_j q$
  \end{tabular}
\end{center}

The one-loop determinants to be discussed, (\ref{SUSY-sphere-flux-reg}) and (\ref{A-omega-reg}), in SUSY localization are regularized if we impose the linear constraints
\begin{equation} \label{PV-linear}
\sum_J \epsilon_J =\sum_J \epsilon_J  a_J = \sum_J \epsilon_J b_J = \sum_J \epsilon_J c_J =0\,.
\end{equation}
An analysis with the heat kernel or the explicit enumeration of eigenvalues shows that we can regularize the bosonic and fermionic determinants independently if we impose the quadratic conditions%
\footnote{%
We can for example take $N_\text{PV}=3$, $(\epsilon_1,\ldots,\epsilon_5)=(+1,+1,-1,-1,-1)$, $b_j= c_j= 1$ for all $j$, and $(a_1,\ldots,a_5)=(3,3,1,1,4)$.
}
\begin{equation} \label{PV-quadratic}
\sum_J \epsilon_J a_J^2 = \sum_J \epsilon_J b_J^2 = \sum_J \epsilon_J c_J^2 = \sum_J \epsilon_J a_J b_J = \sum_J \epsilon_J b_J c_J = \sum_J \epsilon_J c_J a_J=0 \,.
\end{equation}

For later convenience we define the following combinations of  parameters.
\begin{equation}
 C_0 := \prod_j |a_j|^{-\epsilon_j} \,,   
\end{equation}
\begin{equation}
C_1 := \sum_j \epsilon_j b_j \log |a_j|\,, 
\qquad
C_2 := \sum_j \epsilon_j a_j \log |a_j|\,, 
\qquad
C_3 := \sum_j \epsilon_j c_j \log| a_j|\,.
\end{equation}
\begin{equation}
\Xi_1 := \sum_j \epsilon_j b_j \text{sgn}(a_j)\,, \quad
\Xi_2 := \sum_j \epsilon_j |a_j| \,, 
\end{equation}
\begin{equation} \label{def-Xi-34}
\Xi_3 := \sum_j \epsilon_j c_j \text{sgn}(a_j)\,,
\quad
\Xi_4 := \sum_j \epsilon_j  \text{sgn}(a_j)\,.
\end{equation}

\subsection{A counterterm action for the chiral multiplet}
\label{sec:univ-count}

At the classical level, a field theory is specified by the action functional.
To define a theory at the quantum level, one must specify the counterterms in addition to a regularization method.
 We now propose a counterterm action $S_\text{ct}$ which, combined with the Pauli-Villars regularization above, defines unambiguously the theory of a chiral multiplet in the background of gauge and gravity multiplets.
Let us denote the renormalized action for the background fields by $S_\text{ren}$ and the regularized one-loop determinant by $Z_\text{1-loop}^\text{PV}$.
The partition function, which is a functional of the gravity multiplet and the $U(1)_\text{F}$ gauge multiplet, is given as
\begin{equation} \label{Z-functional}
  Z[g_{\mu\nu},\sigma,v_\mu,\ldots, \text{fermions}] = \lim_{\Lambda\rightarrow +\infty}  e^{-S_\text{ren} - S_\text{ct}} Z_\text{1-loop}^{\rm PV} \,.
\end{equation}
The limit $\Lambda \rightarrow + \infty$ is taken pointwise when $\Lambda$ is not constant in spacetime.

We will give the counterterm action $S_\text{ct}$ in terms of a twisted superpotential $\widetilde{W}_\text{ct}$.
As is well known a twisted chiral multiplet can be formed from a gauge multiplet.
The scalars $\sigma$ and $\Lambda$ are the bottom components of the $U(1)_\text{F}$ and $\widetilde{U(1)}_\text{PV}$ gauge multiplets.
As reviewed in Appendix \ref{sec:relat-betw-mult} the gravity multiplet also gives rise to a twisted chiral multiplet%
\footnote{%
The gravity multiplet contains a gauge field $A^\text{R}_\mu$ for the vector R-symmetry, and one can construct the corresponding gauge multiplet from the fields in the gravity multiplet.
The R-symmetry gauge multiplet then gives rise to a twisted chiral multiplet whose scalar component is $\widehat{\mathcal{H}}$ (up to a multiplicative numerical factor).
$\widehat{\mathcal{H}}$ differs from the graviphoton field strength $\mathcal{H}$ by terms involving gravitini.
}
whose scalar component is denoted as $\widehat{\mathcal{H}}$.

Let $\mu>0$ be an arbitrary (renormalization) mass scale.
Our counterterm action $S_\text{ct}$ is constructed from the twisted superpotential
\begin{equation}\label{Wt-ct}
\begin{aligned}
  \widetilde{W}_\text{ct} (\sigma,\widehat{\mathcal{H}},\Lambda) 
&=
- \frac{\widehat{\mathcal{H}}   }{8\pi} 
\sum_j \epsilon_j \log \frac{i a_j \Lambda}{\mu}
\\
&\qquad
+ \frac{1}{ 4\pi} \sum_j\epsilon_j(a_j \Lambda + b_j
\sigma + \frac{ c_j q}{2} \widehat{\mathcal{H}})\log \frac{a_j
  \Lambda + b_j \sigma + \frac{ c_j q}{2} \widehat{\mathcal{H}}}{\mu\, e }
\end{aligned}
\end{equation}
by applying the rule (\ref{twisted-F-term-coupling-rigid}).
We choose the branch cut of the logarithm to be along the negative real axis.
In later sections we demonstrate that the counterterm action given by
(\ref{Wt-ct-expanded}) successfully regularizes the one-loop
determinants obtained by supersymmetric localization.
This is the main result of this note.

The second term in (\ref{Wt-ct}) is essentially minus the familiar effective twisted superpotential from integrating out massive chiral multiplets \cite{Witten:1993yc}; here they are coupled to the gauge multiplets not only for the usual $U(1)_\text{F}$ symmetry, but also for the $\widetilde{U(1)}_\text{PV}$ and R-symmetries.
The requirement of supersymmetry and gauge symmetry does not fix the counterterm completely; in including $i a_j$ in the first term, we made a particular choice for the finite counterterm.

To exhibit more features of  (\ref{twisted-F-term-coupling-rigid}),
we expand it in $1/\Lambda$, assuming that $\Lambda>0$.
\begin{equation}\label{Wt-ct-expanded}
  \begin{aligned}
  \widetilde{W}_\text{ct} (\sigma,\widehat{\mathcal{H}},\Lambda) &=
 \frac{1-q }{8\pi}\widehat{\mathcal{H}}   \log\frac{\Lambda}{\mu}
+\frac{1 }{4 \pi }\left(C_1 -\log \frac{\Lambda }{\mu }+i\frac{\pi } {2} \Xi_1 \right) \sigma 
\\
&\qquad
+ \frac{1}{4 \pi }\left(C_2  + i\frac{\pi  }{2}  \Xi _2  \right) \Lambda
+ \frac{q}{4\pi}\left( C_3  +  i \frac{\pi }{2} \Xi _3   \right) \frac{\widehat{\mathcal{H}} }{2}
\\
&\qquad \qquad
+\frac{1}{4\pi}\left( \log C_0 -i \frac{\pi}{2}\Xi_4 \right) \frac{ \widehat{\mathcal{H}} }{2}
+ \mathcal{O}(\Lambda^{-1})\,.
  \end{aligned}
\end{equation}
The first term of (\ref{Wt-ct-expanded}) gives, via the relation (\ref{twisted-F-term-coupling-rigid}), the supersymmetrization of the well-known counterterm
\begin{equation}
\frac{\hat c_{\rm UV}}{8\pi}\int d^2x \sqrt{g}  \, R \, \log \frac{\Lambda}{\mu} 
\end{equation}
in the action, where $\hat c_{\rm UV}=1-q$ is the UV central charge and $R$ is the scalar curvature.%
\footnote{%
In our convention $R=+2/\ell^2$ for the round sphere of radius $\ell$.
}
The second term describes the renormalization of the FI parameter and the theta angle for $U(1)_\text{F}$; it implies that the renormalized couplings $(r(\mu),\theta)$ and the bare couplings~$(r_0(\Lambda),\theta_0)$ are related as%
\footnote{%
The renormalized FI parameter $r(\mu)$ is the coefficient of  $D/2\pi$ in $\mathcal{L}_\text{ren}$, while the bare parameter $r_0(\Lambda)$ is the corresponding coefficient in $\mathcal{L}_\text{ren}+\mathcal{L}_\text{ct} $, where the action $S$ and the Lagrangian $\mathcal{L}$ are related as $S=\int d^2x \sqrt{g} \mathcal{L}$.
See (\ref{FI}).
}
\begin{equation} \label{FI-renormalization}
\log \frac{\Lambda }{\mu} - C_1  +  r(\mu)= r_0  (\Lambda) \,,
\qquad
\theta  + \frac{\pi}{2} \Xi_1 = \theta_0 \,.
\end{equation}
Similarly the terms in the last two lines renormalize the FI-$\theta$ couplings for $\widetilde{U(1)}_\text{PV}$ and R-symmetry.   See (\ref{R-gauge-from-grav}).

In Appendix \ref{FI-flat} we demonstrate by a flat space computation that the counterterm indeed cancels the divergence in the effective FI-parameter.
We also show that the relation (\ref{FI-renormalization}) gives rise to the standard $\sigma$-dependent effective FI-parameter \cite{Witten:1993yc}.
Using $r(\mu)$ and $\theta$ we define the complexified FI parameter
\begin{equation}
  t = r - i \theta \,.  
\end{equation}
The part of the renormalized action $S_\text{ren}$ that survives in localization calculations below is the FI and the theta terms, both of which can be represented by the twisted superpotential $-t\Sigma/4\pi$.

\section{SUSY sphere partition function}

In this section we apply our renormalization procedure to the two-dimensional sphere partition function \cite{Benini:2012ui,Doroud:2012xw}.
Let us write $\sigma=\sigma_1+i\sigma_2$, $\overline{\sigma}=\sigma_1 -i \sigma_2$.
The supergravity background is given in our convention as
\begin{equation}
  ds^2= f(\theta)^2 d\theta^2 + \ell^2\sin^2\theta d\varphi^2\,,
\qquad 0\leq\theta\leq \pi\,,
\end{equation}
\begin{equation}
{\rm z}=\overline{\rm z}=0  \,,
\quad
A^\text{R}= \frac12\left(1-\frac \ell f\right)d\varphi\,,
\quad
  \mathcal H=\overline{\mathcal H}=-\frac{i}{f}\,,
\end{equation}
\begin{equation} \label{SUSY-sphere-gauge}
v_r=0\,,\qquad
v_\varphi = \frac{B}{2}(1-\cos\theta)\,,
\qquad
V_\varphi=- \frac{q}{2}\left(1-\frac \ell f\right)
\,,
\end{equation}
\begin{equation} \label{SUSY-sphere-gauge2}
\sigma_1 = {\rm const.}\,, \qquad
D= -\frac{i}{f} \sigma_1\,,
\qquad
\sigma_2 = - \frac{B}{2 \ell} \,,
\end{equation}
with vanishing gravitini.
Here $B\in \mathbb{Z}$.
 The function $f(\theta)$ is assumed to sasisfy
\begin{equation}
  f(\pi-\theta)=f(\theta)\,,
\qquad
f(\theta)\simeq \ell+ \mathcal O(\theta^2)
\end{equation}
and is otherwise arbitrary.
For the PV gauge multiplet we have, in analogy with (\ref{SUSY-sphere-gauge2}),
\begin{equation}
D^\text{PV} =  -\frac{i  \Lambda}{f}, \qquad \Lambda = \text{constant}>0\,.
\end{equation}
The SUSY parameters in the convention explained in the appendix are
\begin{equation}
  \begin{pmatrix}
    \epsilon_- \\
    \epsilon_+
  \end{pmatrix}
=
  \begin{pmatrix}
e^{-i\varphi} \sin\frac{\theta}{2}
\\
-\cos\frac{\theta}{2}
  \end{pmatrix}
\,,
\qquad
  \begin{pmatrix}
    \overline\epsilon_- \\
    \overline\epsilon_+
  \end{pmatrix}
=
\begin{pmatrix}
-\cos\frac\theta 2
\\
e^{i\varphi} \sin\frac{\theta}{2}
\end{pmatrix}\,.
\end{equation}

The computations in \cite{Benini:2012ui,Doroud:2012xw,Gomis:2012wy} for a single chiral multiplet give a formal, {\it i.e.}, non-convergent, expression for the one-loop determinant as an infinite product:
\begin{equation} \label{SUSY-sphere-flux-formal}
  Z^\text{SUSY}_\text{1-loop}
\text{ ``$=$'' } \prod_{n=0}^\infty \frac{n+1+ \frac{1}{2}|B|-\hat{\sigma}}{n+ \frac{1}{2} |B|+\hat{\sigma}} \,.
\end{equation}
Here $\hat\sigma=i\ell \sigma_1 + \frac{q}{2}$.
The PV-regularized version is
\begin{equation} \label{SUSY-sphere-flux-reg}
  Z^\text{SUSY}_\text{1-loop, reg}
= \prod_{n=0}^\infty 
\bigg[
\frac{n+1+ \frac{1}{2}|B|-\hat{\sigma}}{n+ \frac{1}{2} |B|+\hat{\sigma} }
\prod_j \bigg( \frac{n+1+ \frac{1}{2}|b_j B|-M_j}{n+ \frac{1}{2} |b_j B|+M_j} \bigg)^{\epsilon_j}
\bigg]
\,,
\end{equation}
where
\begin{equation}
  M_j  \equiv { c_j} \frac{q}{2} + i \ell ( a_j \Lambda + b_j {\rm Re}(\sigma) ) \,.  
  \end{equation}

With the Pauli-Villars ghost contributions, the infinite product (\ref{SUSY-sphere-flux-reg}) is absolutely convergent if the constraints   (\ref{PV-linear}) are satisfied.
We can express (\ref{SUSY-sphere-flux-reg}) in terms of gamma functions using the infinite product formula
\begin{equation}
\Gamma(z+1) = e^{-\gamma z} \prod_{k=1}^\infty \frac{e^{\frac{z}{k}}}{ 1+ \frac{z}{k}} \,.
\end{equation}
We remove the absolute value signs, $|B|\rightarrow B$, by applying the equality $\Gamma(x-\frac{k}{2})/\Gamma(1-x-\frac{k}{2})= (-1)^k \Gamma(x + \frac{k}{2})/\Gamma(1-x + \frac{k}{2})$ for integer $k$.%
\footnote{%
Even if we keep the absolute value symbol the result (\ref{PV-linear}) remains valid.
Note that  $\Xi_1$ is an odd integer.
}
We obtain
\begin{equation} \label{SUSY-one-loop-reg}
  \begin{aligned}
  Z^\text{SUSY}_\text{1-loop, reg}
&=
\frac{\Gamma(\hat{\sigma} + \frac{B}{2})}{\Gamma(1-\hat{\sigma} + \frac{B}{2})}
\prod_j \bigg( \frac{\Gamma(M_j + b_j \frac{B}{2})}{\Gamma(1- M_j + b_j \frac{B}{2})} \bigg)^{\epsilon_j}
\\
&= C_0 e^{i  \frac{\pi}{2} \Xi_1  B} e^{(C_3-C_1)q}e^{2 C_1 \hat{\sigma}}e^{2i C_2 \ell \Lambda} (\ell\Lambda)^{1-2\hat{\sigma}}
\frac{\Gamma(\hat{\sigma} + \frac{B}{2})}{\Gamma(1-\hat{\sigma} + \frac{B}{2})}
\left(
1+ \mathcal{O}(\Lambda^{-1})
\right)
\,,
  \end{aligned}
\end{equation}
where the second line is valid for $\Lambda\gg \ell^{-1}$.
The final partition function for a chiral multiplet is
\begin{equation}\label{Z-sphere-CCC}
  \begin{aligned}
  Z^\text{SUSY} 
&=
 \lim_{\Lambda\rightarrow \infty}e^{-S_\text{ren}-S_\text{ct}}    Z^\text{SUSY}_\text{1-loop, reg} 
\\
&=
    e^{-S_{\rm ren}} 
(\ell\mu)^{1-2\hat{\sigma}} \frac{\Gamma(\hat{\sigma} + \frac{B}{2})}{\Gamma(1-\hat{\sigma} + \frac{B}{2})} 
\\
&=
e^{ - i B \theta } e^{ 4\pi i[ r(\mu) - \frac{1}{2\pi}\log(\ell\mu)]\ell {\rm Re}\,\sigma}
(\ell\mu)^{1-q} \frac{\Gamma(\hat{\sigma} + \frac{B}{2})}{\Gamma(1-\hat{\sigma} + \frac{B}{2})} 
\,.
  \end{aligned}
\end{equation}
Here $\hat\sigma=i\ell\, {\rm Re}(\sigma) + \frac{q}{2}$.%

A convenient choice that gives a simple formula is to take $\mu=1/\ell$.
Then
\begin{equation}\label{Z-sphere-simple}
  Z^\text{SUSY} 
=
e^{ 4\pi i r}e^{ - i B\theta} \frac{\Gamma(\hat{\sigma} + \frac{B}{2})}{\Gamma(1-\hat{\sigma} + \frac{B}{2})} 
\,.
\end{equation}
This is the formula that is often quoted in the literature.

For large mass ${\rm Re}\,\sigma$ we can further integrate out the physical chiral multiplet and shift the theta angle.
The sign of the shift is correlated with the sign of the mass (as in the discussion of topological insulators).

\section{A-twisted theory with omega deformation}

The next example is a chiral multiplet in the omega-deformed A-twisted theory.

Let us consider the two-sphere with metric
\begin{equation}
ds^2 = \Omega^2 |dz|^2 = \Omega(r)^2 ( dr^2 + r^2 d\varphi^2) \,,
\end{equation}
where $z=r e^{i\varphi}$.
Let us define the vector field
\begin{equation}
W:=  \partial_\varphi\,.
\end{equation}
Let us use the hat to indicate frame indices.  Then
\begin{equation}
  W_{\hat z}=-\frac{i\overline z}{2} \Omega\,,
\qquad  
  W_{\hat{\overline z}}=\frac{i z}{2} \Omega\,,
\end{equation}
Following \cite{Closset:2014pda} we consider the SUSY parameters 
\begin{equation}
  \begin{pmatrix}
    \epsilon_- \\
    \epsilon_+
  \end{pmatrix}
=
  \begin{pmatrix}
a W_{\hat z}
\\
1
  \end{pmatrix}
\,,
\qquad
  \begin{pmatrix}
    \overline\epsilon_- \\
    \overline\epsilon_+
  \end{pmatrix}
=
 \begin{pmatrix}
1\\
- a W_{\hat{\overline{z}} }
 \end{pmatrix}
\,,
\end{equation}
where the omega-deformation parameter $a$ is an arbitrary constant.
The gravitino variations
vanish when%
\footnote{%
In our convention $\omega_z= i \omega_z{}^{\hat z}{}_{\hat z}$, $\omega_{\overline z}= - i \omega_{\overline z}{}^{\hat{\overline z}}{}_{\hat{\overline z}}$, where $\omega^a{}_b= dx^\mu\omega_\mu{}^a{}_b $ is the usual spin connection determined by $d e^a+\omega^a{}_b\wedge e^b=0$.
}
\begin{equation}
  A^\text{R}_\mu=\frac{1}{2}\omega_\mu\,,  
\qquad
\mathcal H= \frac{a}{2} \epsilon^{\mu\nu}\partial_\mu W_\nu \,,
\qquad
\overline{\mathcal H}=0\,.
\end{equation}

In order to preserve SUSY, we require that
$ \delta\lambda_\pm =0$, $\delta\overline\lambda_\pm  =0$.
These conditions can be rewritten, with rotational symmetry ($W$-invariance) assumed, as
\begin{equation}\label{omega-locus}
D - \frac{\partial_r v_\varphi}{r\Omega^2} = \frac{1}{r\Omega^2} \partial_r \left( \frac{2\sigma }{ a} \right)\,,
\qquad
D + \frac{\partial_r v_\varphi}{r\Omega^2} = \frac{1}{r\Omega^2} \partial_r \left( \frac{a}{2} r^2\Omega^2 \overline \sigma\right)\,.
\end{equation}
They are solved by
\begin{equation}
  D = \frac{1}{r \Omega^2}\partial_r\left( \frac{\sigma}{a} + \frac{a}{4} r^ 2\Omega^2 \overline{\sigma} \right) \,,  
\qquad
  \partial_r v_\varphi =
\partial_r \left(
- \frac{\sigma}{a} + \frac{a}{4} r^2 \Omega^2 \overline{\sigma}
\right)   
\end{equation}
with $\sigma$ and $\overline{\sigma}$ being arbitrary functions of $r$.

References \cite{Benini:2015noa,Closset:2015rna} found the formal expression for the one-loop determinant
\begin{equation} \label{Omega-A-1-loop}
  Z^\text{A-twist, $\Omega$, sphere}_\text{1-loop}
\text{ ``$=$'' }
\prod_{n=0}^\infty \frac{ {\boldsymbol \sigma} + ( n +1   -\frac{B+q }{2} ) a }{ {\boldsymbol \sigma}  + ( n  +\frac{B+q}{2})a}  
  \,.
\end{equation}
Here the constant ${\boldsymbol \sigma}$ is defined by the relations $\sigma(\text{north pole}(r=0)) = {\boldsymbol \sigma} + \frac{B}{2} a$, $\sigma(\text{south pole}) = {\boldsymbol \sigma} - \frac{B}{2} a$.
We assume that the gauge flux for $\widetilde{U(1)}_\text{PV}$ vanishes, so that the associated quantities $\Lambda$ and ${\boldsymbol \Lambda}$, corresponding to $\sigma$ and ${\boldsymbol \sigma}$ respectively, coincide.
The PV-regularized version of (\ref{Omega-A-1-loop}) is then
\begin{equation} \label{A-omega-reg}
  Z^\text{A-twist, $\Omega$, sphere}_\text{1-loop, reg}
=
\prod_{n=0}^\infty 
\left[
\frac{ {\boldsymbol \sigma} + ( n +1 -\frac{B+q }{2} ) a}{ {\boldsymbol \sigma}  + ( n + \frac{B+q}{2}) a}  
\prod_j
\frac{ {\boldsymbol \sigma}_j + ( n +1 - \frac{b_j B+{ c_j} q }{2} ) a}{ {\boldsymbol \sigma}_j  + ( n  + \frac{b_j B+{ c_j} q}{2}) a}  
\right]
\,,
\end{equation}
where
\begin{equation}
{\boldsymbol \sigma}_j = a_j {\boldsymbol \Lambda} + b_j {\boldsymbol \sigma}\,.
\end{equation}
As in the previous section we rewrite this using gamma functions:
\begin{equation} 
  \begin{aligned}
  Z^\text{$\Omega$-$A$}_\text{1-loop, reg}
&=
\frac{\Gamma(\frac{\boldsymbol \sigma}{a}-1 + \frac{B+q}{2})
}{
\Gamma(\frac{\boldsymbol \sigma}{a} - \frac{B+q}{2})
}
\prod_j \bigg( 
\frac{\Gamma(\frac{{\boldsymbol \sigma}_j}{a}-1 + \frac{b_j B+{ c_j q}}{2})
}{
\Gamma(\frac{{\boldsymbol \sigma}_j}{a} - \frac{b_j B+c_j q}{2})
}
\bigg)^{\epsilon_j} \,.
  \end{aligned}
\end{equation}
Let us assume here that ${\boldsymbol \Lambda}>0$.
We find
\begin{equation}
  \begin{aligned}
  Z^\text{A-twist, $\Omega$, sphere}_\text{1-loop, reg}
&=
C_0  e^{C_1 B + C_3 q+\frac{\pi i}{2}  ( \Xi_1 B  + \Xi_3 q - \Xi_4 ) }
\\
&\qquad
\times
  \left(\frac{\boldsymbol\Lambda }{a}\right)^{1-B-q} 
\frac{\Gamma(\frac{\boldsymbol \sigma}{a}-1 + \frac{B+q}{2})
}{
\Gamma(\frac{\boldsymbol \sigma}{a} - \frac{B+q}{2})
}
(1+\mathcal{O}(1/{\boldsymbol \Lambda})) \,.
  \end{aligned}
\end{equation}

The counterterm action can be evaluated in the background configuration by reducing it to the integral of a total derivative expression.
The resulting on-shell value of the master counterterm is such that
\begin{equation}
 e^{- S_{\rm ct} }=
C_0^{-1} \left( \frac{\mu}{\boldsymbol\Lambda }\right)^{1-q-B} e^{-B  C_1 -C_3 q - \frac{\pi i}{2} (\Xi_1 B + \Xi_3    q  - \Xi_4)} \,.
\end{equation}
The physical renormalized action contributes simply $  e^{- S_{\rm ren}} = e^{ +2\pi r(\mu) B} $.
Putting everything together, we see that the ${\boldsymbol \Lambda}$-dependence of the one-loop partition function drops out as it should:
\begin{equation}
  \begin{aligned}
  Z^\text{A-twist, $\Omega$}_\text{1-loop} &= \lim_{ {\boldsymbol \Lambda} \rightarrow \infty}     e^{- S_{\rm ren}}    e^{- S_{\rm ct} }  Z^\text{A-twist, $\Omega$, sphere}_\text{1-loop, reg}
\\
&=
\left(\frac{\mu}{i a}\right)^{1-q-B} e^{ 2\pi r(\mu) B} 
\frac{\Gamma(\frac{\boldsymbol \sigma}{a}-1 + \frac{B+q}{2})
}{
\Gamma(\frac{\boldsymbol \sigma}{a} - \frac{B+q}{2})
}
\,.
  \end{aligned}
\end{equation}

\section{Discussion}
\label{sec:disc-concl}

In this paper we applied the Pauli-Villars regularization to particular one-loop divergences arising from chiral multiplets in two dimensions.
The Pauli-Villars method has been used in other supersymmetric contexts; see for example \cite{West:1985jx,Gaillard:2016nji} and the references therein.

It may be possible and useful to use the central charges, {\it i.e.}, the charges for the graviphotons, as the large masses for the Pauli-Villars regulator fields.
Such a scheme would be more conventional in the sense that the counterterms will depend on the physical vector and gravity multiplets, but not on the Pauli-Villars vector multiplets we introduced in Section \ref{sec:supersymm-pauli-vill}.
A drawback is that we cannot completely recycle the one-loop computations that already exist in the literature because we need to know the dependence on the central charge ${\rm z}$.
The drawback may not be significant because the dependence on the central charge is expected to be essentially the same as the dependence on the twisted mass.
In particular in the index theory approach, the one-loop determinant is computed from the square of the localizing supercharge, in which the central charge enters in the same way as the twisted mass \cite{Closset:2014pda}.
Another complication is that the graviphotons and the central charge are not completely independent because they must obey flux quantization conditions \cite{Closset:2014pda}.

In what sense are our results supersymmetric?
In the given rigid limit of supergravity, the preserved supersymmetries are restricted and finite dimensional.
Our regularization method and the counterterms are invariant under such rigid supersymmetry transformations.
Our regularization scheme, however, has manifest invariance in somewhat more general circumstances as we now explain.

In principle, our regularizations and counterterms are applicable in general, not necessarily supersymmetric, backgrounds.
Without regularization and countertrerms, integrating out matter supermultiplets produces, in an intermediate step, a cut-off dependent expression%
\footnote{%
This is the right hand side of (\ref{Z-functional}) before taking the limit.
}
 that is manifestly invariant under general supersymmetry transformations.
In general the UV cut-off $\Lambda(x)$ itself transforms and is position dependent.
In the limit $\Lambda(x)\rightarrow +\infty$, we expect to obtain a finite functional (\ref{Z-functional}) for the background gravity and gauge supermultiplets.
The point is that this functional is invariant under the general supergravity transformations of the multiplets.
The result of a localization computation as described in the previous paragraph is a specialization of the functional to the specific supersymmetric background.

\section*{Acknowledgements}

This research is supported in part by  JSPS Grants-in-Aid for Scientific Research No.~25287049 and No.~16K05312.
The author acknowledges the Galileo Galilei Institute for Theoretical Physics for providing the productive atmosphere while part of this work was done.
\appendix

\section{Rigid $\mathcal N=(2,2)$ SUSY in curved spacetime}
\label{sec:review-mathcal-n=2}

In \cite{Closset:2014pda}, quantum field theories coupled to a version of two-dimensional $\mathcal N=(2,2)$ supergravity were studied.
The version of supergravity coincides with the dimensional reduction of four-dimensional $\mathcal N=1$ new minimal supergravity, and as such it requires the presence of a $U(1)$ R-symmetry.
We adapt the convention such that it is the vector R-symmetry.
The formulas in this section are for rigid SUSY: they are obtained from those of Appendix \ref{sec:full-supergr-with} by setting gravitini, their variations, $e^{\hat{z}}{}_{\overline{z}}$, and $e^{\hat{\overline{z}}}{}_z$ to zero.

\subsection{SUSY transformations}
\label{sec:susy-transformations}

The SUSY transformations are related to the supercharges as
\begin{equation}
\delta = i\epsilon_+ Q_- - i\epsilon_- Q_+  -i \overline\epsilon_+ \overline Q_- + i\overline \epsilon_- \overline Q_+\,.
\end{equation}
From \cite{Closset:2014pda} and in the convention of \cite{MR2003030},
we have the following supersymmetry transformations for a vector multiplet in the Wess-Zumino gauge:
\begin{equation} \label{SUSY-vector}
  \begin{aligned}
\delta v_{\hat z} & = - \frac{ i }{ 2 } \epsilon_- \overline{\lambda}_-  - \frac{ i }{ 2 } \overline{\epsilon}_- \lambda_- 
\,, 
\qquad
\delta v_{\hat{\overline z}}
 = + \frac{ i }{ 2 } \epsilon_+ \overline{\lambda}_+  + \frac{ i }{ 2 } \overline{\epsilon}_+ \lambda_+ \,,
\\
\delta \sigma & = - i \epsilon_- \overline{\lambda}_+ - i \overline{\epsilon}_{+} \lambda_- \,, \qquad
\quad \delta \overline{\sigma} 
 = - i \overline{\epsilon}_- \lambda_+ - i \epsilon_+ \overline{\lambda}_- \,,
\\
\delta\lambda_-&=+ i\epsilon_- \left( 
D+2i v_{\hat z \hat{\overline{z}}} -\overline{\mathcal H}\sigma-\frac12 [\sigma,\overline\sigma]
\right)
-2\epsilon_+ D_{\hat z}\sigma\,,
\\
    \delta\lambda_+&= +i\epsilon_+ \left( 
D -2i v_{\hat z \hat{\overline{z}}} - \mathcal H\overline\sigma + \frac12 [\sigma,\overline\sigma]
\right)
+ 2\epsilon_- D_{\hat{\overline z}}\overline\sigma\,,
\\
    \delta\overline\lambda_-&=- i\overline\epsilon_- \left( D - 2i v_{\hat z \hat{\overline{z}}} -\mathcal H \overline\sigma-\frac12 [\sigma,\overline\sigma]
\right)
-2\overline\epsilon_+ D_{\hat z}\overline\sigma\,, \qquad
\\
    \delta\overline\lambda_+&= - i\overline\epsilon_+ \left(  D + 2i v_{\hat z \hat{\overline{z}}} - \overline{\mathcal H}\sigma + \frac12 [\sigma,\overline\sigma]
\right) + 2\overline \epsilon_- D_{\hat{\overline z}}\sigma\,,
\\
\delta D & =D_{\hat z}( \overline{\epsilon}_+ \lambda_+ - \epsilon_+ \overline{\lambda}_+ ) - D_{\hat {\overline{z}}}( \overline{\epsilon}_-\lambda_- - \epsilon_- \overline{\lambda}_-) 
\\
&\qquad \qquad
+\frac{i}{2} [\sigma, \epsilon_+ \overline{\lambda}_--\overline{\epsilon}_- \lambda_+] - \frac{i}{2} [\overline{\sigma}, \overline{\epsilon}_+ \lambda_- - \epsilon_- \overline{\lambda}_+]\,. 
  \end{aligned}
\end{equation}

For chiral and anti-chiral multiplets the transformations are
\begin{equation}
  \begin{aligned}
\delta\phi&=   + \epsilon_+\psi_--\epsilon_-\psi_+\,,
\qquad\delta\overline\phi
= -\overline   \epsilon_+\overline \psi_- + \overline \epsilon_- \overline \psi_+\,,
\\
\delta\psi_-&=\epsilon_- F- \overline{\epsilon}_-\left({\rm z}-\sigma -\frac{q}{2}\mathcal H\right)\phi +2i\overline{\epsilon}_+ D_{\hat{z}}\phi\,,
\\
\delta\psi_+&=\epsilon_+ F + \overline{\epsilon}_+\left(\overline{\rm z}-\overline\sigma -\frac{q}{2}\overline{\mathcal H}\right)\phi +2i\overline{\epsilon}_- D_{\hat{\overline z}}\phi\,,
\\
\delta\overline\psi_-&=\overline \epsilon_- \overline F - \epsilon_-  \overline\phi  \left(\overline{\rm z} -\overline \sigma -\frac{q}{2}\overline{\mathcal H}\right) - 2i \epsilon_+ D_{\hat{ z}}\overline \phi\,,
\\
\delta\overline\psi_+&=\overline \epsilon_+ \overline F + \epsilon_+ \overline\phi \left({\rm z} - \sigma -\frac{q}{2}{\mathcal H}\right) - 2i \epsilon_- D_{\hat{\overline z}}\overline \phi\,,
\\
\delta F&=-\left(\overline{\rm z}-\overline\sigma-\frac{q}{2}\overline{\mathcal H}\right) \overline\epsilon_+\psi_- - \left({\rm z}-\sigma-\frac{q}{2}{\mathcal H}\right)\overline{\epsilon}_- \psi_+ 
\\
&\qquad\qquad
+ 2i  \overline{\epsilon}_+ D_{\hat z}\psi_+ - 2i \overline{\epsilon}_-  D_{\hat{\overline z}}\psi_- -i (\overline{\epsilon}_+ \overline{\lambda}_--\overline{\epsilon}_- \overline{\lambda}_+)\phi\,.
\\
\delta \overline{F} &= +\epsilon_+ \overline\psi_-  \left ({\rm z}-\sigma-\frac{q}{2}\mathcal H\right) + \epsilon_-\overline\psi_+\left (\overline{\rm z}- \overline\sigma - \frac{q}{2}\overline{\mathcal H}\right)
\\
&\qquad \qquad
 +2i  \epsilon_+ D_{\hat z}\overline\psi_+ - 2i  \epsilon_- D_{\hat{\overline z}}\overline{\psi}_- -i\overline\phi (\epsilon_+\lambda_- - \epsilon_- \lambda_+)\,.
  \end{aligned}
\end{equation}

Similarly, 
fields in twisted chiral and anti-chiral multiplets transform as
\begin{equation}
  \begin{aligned}
&
 \qquad \qquad \qquad
\delta v = \overline{\epsilon}_+ \chi_- -\epsilon_- \overline{\chi}_+\,, \\
&\delta\chi_- = 2i \epsilon_+ D_{\hat{z}} v +\epsilon_- E\,,
\qquad
\delta \overline{\chi}_+ 
= 2i \overline{\epsilon}_- D_{\hat{\overline{z}}}v + \overline{\epsilon}_+ E\,,\\
&
\qquad \qquad
\delta E = 2i \epsilon_+ D_{\hat z} \overline{\chi}_+ -2i \overline{\epsilon}_- D_{\hat{\overline{z}}} \chi_-\,.
  \end{aligned}
\end{equation}
\begin{equation}
  \begin{aligned}
&\ \  \qquad\qquad \qquad  \delta \overline{v}  = - \epsilon_+ \overline{\chi}_- + \overline{\epsilon}_- \chi_+ \,, \\
&  \delta \overline{\chi}_-  = - 2i \overline{\epsilon}_+ D_{\hat{z}} \overline{v} + \overline{\epsilon}_- \overline{E} \,, 
\qquad
  \delta \chi_+  = - 2i \epsilon_- D_{\hat{\overline{z}}} \overline{v} + \epsilon_+ \overline{E} \,, \\
&\qquad\qquad \qquad  \delta \overline{E}  = 2i \overline{\epsilon}_+ D_{\hat{z}} \chi_+ - 2i \epsilon_- D_{\hat{\overline{z}}} \overline{\chi}_- \,.
  \end{aligned}
  \end{equation}

\subsection{Relations between multiplets}
\label{sec:relat-betw-mult-rigid}

An abelian gauge multiplet gives rise to twisted chiral and twisted anti-chiral multiplets, indicated by superscript ``ga'':
\begin{equation}
  \begin{aligned}
& v^{\rm ga} = \sigma \,, \qquad  \chi^{\rm ga}_- = -i \lambda_- \,, \qquad \overline{\chi}^{\rm ga}_+ =  i \overline{\lambda}_+ \,, \\
& \qquad E^{\rm ga} =   D+2i v_{\hat z \hat{\overline{z}}} -\overline{\mathcal H}\sigma
\,.
  \end{aligned}
\end{equation}
\begin{equation}
  \begin{aligned}
&    \overline{v}^{\rm ga}  = \overline{\sigma} \,, \qquad  \overline{\chi}^{\rm ga}_- = i \overline{\lambda}_- \,, \qquad \chi^{\rm ga}_+ = - i \lambda_+ \,, \\
&\qquad \overline{E}^{\rm ga} = D - 2i v_{\hat z \hat{\overline{z}}} -\mathcal H \overline\sigma
\,.
  \end{aligned}
\end{equation}

Holomorphic and anti-holomorphic functions $\widetilde{W}(v^{(j)})$ and $\overline{\widetilde{W}}(\overline{v}^{(j)})$ give rise to twisted chiral and twisted anti-chiral multiplets

\begin{equation}
  \begin{aligned}
&  v^{\widetilde{W}} :=  \widetilde{W}\,,
\qquad
  \chi^{\widetilde{W}}_- :=  \chi^{(j)}_- \partial_j \widetilde{W}\,,
\qquad
  \overline{\chi}^{\widetilde{W}}_+ :=  \overline{\chi}^{(j)}_+ \partial_j \widetilde{W}\,,
\\
&
\qquad\qquad
  E^{\widetilde{W}} :=  E^{(j)} \partial_j \widetilde{W} + \chi^{(i)}_- \overline{\chi}^{(j)}_+ \partial_i \partial_j \widetilde{W} \,.
  \end{aligned}
\end{equation}

\subsection{Supersymmetric actions}
\label{sec:supersymm-acti}

\subsubsection{Chiral multiplet}
\label{sec:chiral-multiplet}

For the chiral multiplet, the supersymmetric UV action is
\begin{equation} \label{action-chiral}
  \begin{aligned}
S_\text{chi}
&=\int d^2x \sqrt{g}\Bigg[
2 D_{\hat{z}} \overline{\phi} D_{\hat{\overline{z}}} \phi + 2 D_{\hat{\overline{z}}}  \overline{\phi} D_{\hat{z}} \phi   
 - \overline{\phi} D\phi +\overline{\phi} \left( \frac{q}{4}R + \frac{\overline{\rm z}  }{2} \mathcal{H} + \frac{ {\rm z} }{2} \overline{\mathcal{H}} \right) \phi
\\
&\qquad
+ \overline{\phi} ( \overline{\sigma} - \overline{\rm z} + \frac{q}{2} \overline{\mathcal{H}}) (  \sigma -{\rm z} + \frac{q}{2} \mathcal{H}) \phi
+ 2i \overline{\psi}_+ D_{\hat{z}} \psi_+ - 2i \overline{\psi}_- D_{\hat{\overline{z}}} \psi_-
\\
&\qquad
+ \overline{\psi}_+ \left(\overline{\sigma}-\overline{\rm z} + \frac{q}{2}\overline{\mathcal{H}}\right) \psi_-
+ \overline{\psi}_- \left( \sigma - {\rm z} + \frac{q}{2} \mathcal{H} \right)\psi_+
\\
&\qquad
+i \left(\overline{\psi}_- \overline{\lambda}_+ -  \overline{\psi}_+\overline{\lambda}_- \right)\phi
+i \overline{\phi} \left( \lambda_-\psi_+ - \lambda_+ \psi_- \right) - \overline{F} F
\Bigg]
  \end{aligned}
\end{equation}
Here
\begin{equation}
  D_{\mu}=\nabla_{\mu} + i v_\mu + i q A^\text{R}_{\mu} + \frac{\rm z}{2} \overline{C}_\mu - \frac{\overline{\rm z}}{2} C_\mu \,.
\end{equation}

The superpotential coupling for several chiral multiplets $(\phi^i,\psi_\mp^i,F^i)$ is
\begin{equation}
  S =   - \int d^2x \sqrt{g} \left[
    F^j \partial_i W + \psi^i_- \psi_+^j \partial_i\partial_j W + \overline{F}^j \partial_i \overline{W} - \overline{\psi}^i_- \overline{\psi}_+^j \partial_i\partial_j \overline{W}
\right]\,.
\end{equation}

\subsubsection{Vector multiplet}
\label{sec:vector-multiplet}

For the vector multiplet, the action is
\begin{equation}
  \begin{aligned}
  S_\text{vec} & = \int d^2x \sqrt{g} \, {\rm Tr}\, \Bigg[
D_{\hat{\overline{z}}}\overline{\sigma} D_{\hat z}\sigma + D_{\hat z}\overline{\sigma} D_{\hat{\overline{z}}} \sigma + \frac{1}{8} [ \sigma, \overline{\sigma} ]^2
\\
&\qquad
+ \frac{1}{2} \left(- 2i v_{\hat{z}\hat{\overline{z}}} +\frac{1}{2} \overline{\mathcal{H}}\sigma - \frac{1}{2} \mathcal{H} \overline{\sigma} \right)^2
-\frac{1}{2} \left( - D + \frac{1}{2} \overline{\mathcal{H}}\sigma + \frac{1}{2} \mathcal{H} \overline{\sigma} \right)^2
\\
&\qquad\qquad
+ i \overline{\lambda}_+ D_{\hat z} \lambda_+ - i \overline{\lambda}_- D_{\hat{\overline{z}}} \lambda_- 
+ \frac{1}{2} \overline{\lambda}_- [ \sigma, \lambda_+] 
+ \frac{1}{2} \overline{\lambda}_+ [ \overline{\sigma}, \lambda_- ] 
\Bigg]  \,.
  \end{aligned}
\end{equation}

We also have the FI-term
\begin{equation} \label{FI}
S_\text{FI} =   r \int d^2 x \sqrt{ g } D 
\end{equation}
and the $\theta$-term
\begin{equation}
  S_\theta = i \frac{\theta}{2\pi} \int dv\,.  
\end{equation}

\subsubsection{Twisted chiral multiplet}
\label{sec:twist-chir-mult-rigid}

Given a twisted superpotential $\widetilde{W}(v^j)$ for twisted chiral multiplets $(v^j, \chi^j_-, \overline{\chi}^j_+, E^j)$, the corresponding action is given as
\begin{equation} \label{twisted-F-term-coupling-rigid}
S = -  \int d^2x \sqrt{g}\Big( E^j\partial_j \widetilde{W}  + \chi^i_- \overline{\chi}^j_+ \partial_i \partial_j \widetilde{W}    + \overline{\mathcal{H}}\widetilde{W} + \overline{E}^j \partial_j \overline{\widetilde{W}}   - \overline{\chi}^i_- \chi_+^j \partial_i \partial_j \overline{\widetilde{W}}   + \mathcal{H} \overline{\widetilde{W}} \Big)  \,.
\end{equation}

The FI and theta terms above arise from the twisted superpotential
\begin{equation}
\widetilde{W} = - \frac{t\Sigma}{4\pi} \,.
\end{equation}

\section{Full $\mathcal N=(2,2)$  supergravity}
\label{sec:full-supergr-with}

Earlier papers include \cite{Howe:1987ba,Grisaru:1994dm,Grisaru:1995dr,Gates:1995du,Ketov:1996es}.
\subsection{Supergravity transformations}
\label{sec:supergr-transf}

The symbol $D_\mu$ denotes the minimal derivative that is covariant with respect to general coordinate, local Lorentz, R-symmetry, and central charge transformations.
As such it is constructed from the partial derivative and the appropriate gauge fields only.

In describing supergravity, we often use complex coordinates $(x^\mu)=(z,\overline{z})$ and the corresponding frame one-forms (vielbein) $(e^a)=(e^{\hat{z}},e^{\hat{\overline{z}}})$.
Thus the Greek indices $\mu,\nu,\ldots$ are for coordinates, and the Latin indices $a,b,\ldots$ are for the frame (or flat, or tangent space) indices.
The latter take a value $\hat{z}$ or $\hat{\overline{z}}$.
The metric takes the form 
\begin{equation}
  ds^2= g_{\mu\nu}dx^\mu dx^\nu = \eta_{ab} e^a e^b \,,
\end{equation}
\begin{equation}
(\eta_{ab}) =  
\begin{pmatrix}
  0 & 1/2 \\
  1/2 & 0
\end{pmatrix}
\,,
\qquad
e^a= e^a{}_\mu dx^\mu \,.
\end{equation}

\subsubsection{Gravity multiplet}
\label{sec:gravity-multiplet}

We work with $\mathcal{N}=(2,2)$ $U(1)_V$ supergravity.
The vielbein and the graviphotons transform as
\begin{equation}
  \begin{aligned}
&    \delta e^{\hat{z}}{}_z  = + 2 \overline{\epsilon}_+ \psi_{+ z} + 2 \epsilon_+ \overline{\psi}_{+ z}  \,,  
\qquad
    \delta e^{\hat{z}}{}_{\overline{z}}  =  + 2 \overline{\epsilon }_+ \psi _{+ \overline{z}}+2 \epsilon _+    \overline{\psi }_{+\overline{z}} \,,
\\
 &   \delta e^{\hat{\overline{z}}}{}_{z}  =  -2 \overline{\epsilon }_- \psi_{- z}-2 \epsilon _- \overline{\psi   }_{-z} \,, 
\qquad
    \delta e^{\hat{\overline{z}}}{}_{\overline{z}}  =  -2 \overline{\epsilon }_- \psi _{- \overline{z}}-2 \epsilon _-    \overline{\psi }_{- \overline{z}}\,,
  \end{aligned}
\end{equation}
\begin{equation}
  \begin{aligned}
&    \delta C_z = + 2 i \overline{\epsilon }_+ \psi _{- z} +2 i \epsilon _-    \overline{\psi }_{+ z} \,,
\qquad
\delta  C_{\overline{z}} = + 2 i \epsilon_-\overline{\psi}_{+\overline{z}}+2 i \overline{\epsilon}_+\psi_{- \overline{z}}\,,  
\\
&\delta  \overline{C}_z = -2 i \overline{\epsilon}_-\psi_{+z}-2 i \epsilon_+\overline{\psi}_{-z}\,, 
\qquad
\delta  \overline{C}_{\overline{z}} = -2 i \overline{\epsilon}_-\psi_{+ \overline{z}}-2 i \epsilon_+\overline{\psi}_{- \overline{z}}\,.
  \end{aligned}
\end{equation}

Using the gamma matrices 
\begin{equation}
\gamma_{\hat z} =
\begin{pmatrix}
  & 0 \\
-1 &
\end{pmatrix}
\,,\qquad
\gamma_{\hat{\overline{z}}} =
\begin{pmatrix}
  & -1 \\
0 &
\end{pmatrix} \,,
\qquad
\gamma = 
\begin{pmatrix}
  +1 & 0 \\
  0& -1
\end{pmatrix}
\end{equation}
and the conventions
\begin{equation}
(\varepsilon_{\alpha\beta})=
\begin{pmatrix}
&  +1 \\
-1 &
\end{pmatrix}
\,,
\qquad
(\epsilon_\alpha) = 
\begin{pmatrix}
\epsilon_- 	\\
\epsilon_+
\end{pmatrix}
\,,
\qquad
\epsilon \gamma^a \overline{\psi}_\mu = \varepsilon^{\alpha\beta} \epsilon_\beta \gamma^a{}_\alpha{}^\gamma \overline{\psi}_\mu{}_\gamma
\qquad
\text{etc.,}
\end{equation}
we can also write
\begin{equation}
  \delta e^a{}_\mu = - \overline{\epsilon}\gamma^a \psi_\mu - \epsilon \gamma^a \overline{\psi}_\mu\,,  
\end{equation}
\begin{equation}
  \delta C_\mu = i \overline{\epsilon}(1+\gamma) \psi_\mu - i \epsilon (1- \gamma) \overline{\psi}_\mu\,,  
\qquad
  \delta \overline{C}_\mu = i \overline{\epsilon}(1 - \gamma) \psi_\mu - i \epsilon (1 + \gamma) \overline{\psi}_\mu\,.
\end{equation}
For the gravitini and the R-symmetry gauge field, we have
\begin{equation}\label{delta-psi-eta}
\delta \psi_\mu = i \mathcal{D}_\mu\epsilon \,,
\qquad
\delta \overline{\psi}_\mu = i \overline{\mathcal{D}}_\mu\overline{\epsilon}  \,,
\end{equation}
\begin{equation}\label{delta-AR-eta}
  \delta A^{\rm R}_\mu = \frac{1}{2} \overline{\epsilon}\gamma_\mu\gamma \epsilon^{\rho\sigma} \mathcal{D}_\rho \psi_\sigma - \frac{1}{2} \epsilon \gamma_\mu\gamma \epsilon^{\rho\sigma} \overline{\mathcal{D}}_\rho \overline{\psi}_\sigma 
 +i \psi_\mu\overline{\eta} -i \overline{\psi}_\mu \eta\,.
\end{equation}
We introduced the following notations.
\begin{equation} \label{cal-D-epsilon}
\mathcal{D}_\mu\epsilon = \left(\partial_\mu + i (\omega+K)_\mu \frac{\gamma}{2} + i A^{\rm R}_\mu \right) \epsilon  +\gamma_\mu\eta \,,
\end{equation}
\begin{equation}\label{cal-Dbar-epsilon}
\overline{\mathcal{D}}_\mu\overline{\epsilon} = \left(\partial_\mu + i (\omega+K)_\mu \frac{\gamma}{2} -i A^{\rm R}_\mu \right)  \overline\epsilon  +\gamma_\mu\overline{\eta} \,,
\end{equation}
\begin{equation}
  K_\mu := \overline{\psi}_\mu \gamma^\nu \gamma\psi_\nu   - \overline{\psi}_\nu \gamma^\nu \gamma\psi_\mu  \,,
\end{equation}
\begin{equation}
 \mathcal H = -i \epsilon^{\mu\nu} \partial_\mu C_\nu \,, \qquad \overline{\mathcal H} = -i \epsilon^{\mu\nu} \partial_\mu \overline C_\nu  \,,
\end{equation}
\begin{equation} \label{eta-def}
\eta: = - \frac{ i}{4} \widehat{\mathcal{H}} (1-\gamma)\epsilon + \frac{ i}{4} \widehat{\overline{\mathcal{H}}} (1 + \gamma)\epsilon \,,
\end{equation}
\begin{equation} \label{etabar-def}
\overline{\eta} := - \frac{ i}{4} \widehat{\mathcal{H}}  (1+\gamma)\overline{\epsilon} + \frac{ i}{4} \widehat{\overline{\mathcal{H}}} (1 - \gamma)\overline{\epsilon}  \,,
  \end{equation}
  \begin{equation} \label{H-hat-H-bar-hat}
   \widehat{\mathcal{H}} := \mathcal{H} + i \epsilon^{\mu\nu} \overline{\psi}_\mu(1+\gamma) \psi_\nu  
\,,
\qquad
   \widehat{\overline{\mathcal{H}}} := \overline{\mathcal{H}} + i \epsilon^{\mu\nu} \overline{\psi}_\mu(1-\gamma) \psi_\nu   \,.
  \end{equation}
The equations (\ref{cal-D-epsilon}) and (\ref{eta-def}) define a covariant derivative $\mathcal{D}_\mu$, which acts on a general spinor with R-charge $+1$ and appears in other formulas such as (\ref{delta-AR-eta}).
Similarly~$\overline{\mathcal{D}}_\mu$ defined by (\ref{cal-Dbar-epsilon}) and (\ref{etabar-def}) acts on a spinor with R-charge $-1$.
When acting on $\psi_\sigma$ and $\overline\psi_\sigma$, the covariant
derivatives $\mathcal{D}_\rho$ and $\overline{\mathcal{D}}_\rho$ are defined so that they do not contain Christoffel symbols.%
\footnote{%
The same is true for the four-dimensional new minimal supergravity \cite{Sohnius:1981tp}. 
If we included the torsionless Christoffel symbol constructed from the metric (or the vielbein) in $\mathcal{D}_\rho$ and $\overline{\mathcal{D}}_\rho$ they would cancel out in $\epsilon^{\rho\sigma}\mathcal{D}_{\rho} \psi_{\sigma}$.
Thus the formula (\ref{delta-AR-eta}) is generally covariant.
If we included the Christoffel symbol corresponding to $\omega_\mu+K_\mu$ which has torsion, it would contribute, but this would be the wrong definition.
}

\subsubsection{Gauge multiplet}
\label{sec: sugra-gauge-multiplet}

\begin{equation} \label{eq:sugra-gauge-multiplet} 
  \begin{aligned}
\delta v_{\hat z} & =- \frac{ i }{ 2 } \epsilon_- \overline{\lambda}_-  - \frac{ i }{ 2 } \overline{\epsilon}_- \lambda_- + \overline{\sigma} \overline{\epsilon}_+\psi_{-\hat z} + \sigma \overline{\epsilon}_-\psi_{+\hat z}+ \sigma \epsilon_+ \overline{\psi}_{-\hat z} + \overline{\sigma} \epsilon_- \overline{\psi}_{+\hat z} 
\,, 
\\
\delta v_{\hat{\overline z}}
 &
 = + \frac{ i }{ 2 } \epsilon_+ \overline{\lambda}_+  + \frac{ i }{ 2 } \overline{\epsilon}_+ \lambda_+ + \overline{\sigma} \overline{\epsilon}_+\psi_{-\hat{\overline z}} + \sigma \overline{\epsilon}_-\psi_{+\hat{\overline z}}+ \sigma \epsilon_+ \overline{\psi}_{-\hat{\overline z}} + \overline{\sigma} \epsilon_- \overline{\psi}_{+\hat{\overline z}}  \,,
\\
\delta \sigma & = - i \epsilon_- \overline{\lambda}_+ - i \overline{\epsilon}_{+} \lambda_- \,, 
\\
 \delta \overline{\sigma} & = - i \overline{\epsilon}_- \lambda_+ - i \epsilon_+ \overline{\lambda}_- \,,
\\
\delta\lambda_-&=+ i\epsilon_- \left( 
D+2i \widehat{v}_{\hat z \hat{\overline{z}}} -\widehat{\overline{\mathcal H}}\sigma-\frac12 [\sigma,\overline\sigma] \right) -2\epsilon_+ \mathcal{D}^\text{SG}_{\hat z}\sigma\,,
\\
    \delta\lambda_+&= +i\epsilon_+ \left( 
D -2i \widehat{v}_{\hat z \hat{\overline{z}}} - \widehat{\mathcal H}\overline\sigma + \frac12 [\sigma,\overline\sigma]
\right)
+ 2\epsilon_- \mathcal{D}^\text{SG}_{\hat{\overline z}}\overline\sigma\,,
\\
    \delta\overline\lambda_-&=- i\overline\epsilon_- \left( D - 2i \widehat{v}_{\hat z \hat{\overline{z}}} -\widehat{\mathcal H} \overline\sigma-\frac12 [\sigma,\overline\sigma]
\right)
-2\overline\epsilon_+ \mathcal{D}^\text{SG}_{\hat z}\overline\sigma\,, 
\\
    \delta\overline\lambda_+&= - i\overline\epsilon_+ \left(  D + 2i \widehat{v}_{\hat z \hat{\overline{z}}} - \widehat{\overline{\mathcal H}}\sigma + \frac12 [\sigma,\overline\sigma]
\right) + 2\overline \epsilon_- \mathcal{D}^\text{SG}_{\hat{\overline z}}\sigma\,,
  \end{aligned}
\end{equation}
\begin{equation}
  \begin{aligned}
\delta D & = \overline{\epsilon}_+ D_{\hat z}\lambda_+ - \epsilon_+ D_{\hat z}\overline{\lambda}_+  -  \overline{\epsilon}_-D_{\hat {\overline{z}}}\lambda_- + \epsilon_- D_{\hat {\overline{z}}}\overline{\lambda}_-
\\
&\quad
+\frac{i}{2} [\sigma, \epsilon_+ \overline{\lambda}_--\overline{\epsilon}_- \lambda_+] - \frac{i}{2} [\overline{\sigma}, \overline{\epsilon}_+ \lambda_- - \epsilon_- \overline{\lambda}_+]
\\
&\quad \quad 
-\frac{i}{2}  \widehat{\mathcal H} \overline{\epsilon}_- \lambda_+ -\frac{i}{2}  \widehat{\overline{\mathcal H}} \epsilon_- \overline{\lambda}_+ -\frac{i}{2} \widehat{\overline{\mathcal H}} \overline{\epsilon}_+ \lambda_- -\frac{i}{2}  \widehat{\mathcal H}\epsilon_+ \overline{\lambda}_- 
\\
&\quad \quad \quad
+ \epsilon^{\mu\nu} (\overline{\sigma} \epsilon_- \mathcal{D}_\mu \overline{\psi}_{+\nu} -\sigma  \epsilon_+\mathcal{D}_\mu \overline{\psi}_{-\nu} - \sigma \overline{\epsilon}_- \mathcal{D}_\mu \psi_{+\nu} - \overline{\sigma}  \overline{\epsilon}_+\mathcal{D}_\mu \psi_{-\nu}  )
\,. 
  \end{aligned}
\end{equation}

We defined the SUSY covariant gauge field strength
\begin{equation} \label{gauge-field-srength-SUSY-covariant}
  \begin{aligned}
  \widehat{v}_{\hat{z}\hat{\overline{z}}} &:=
v_{\hat{z}\hat{\overline{z}}}  - \frac{1}{2}\left(\psi_{+\hat{z}}\overline{\lambda}_+ + \overline{\psi}_{+\hat{z}} \lambda_+ + \overline{\psi}_{-\hat{\overline{z}}} \lambda_- + \psi_{-\hat{\overline{z}}}\overline{\lambda}_- \right) 
\\
&\qquad
- \frac{1}{2}\overline{\sigma} \epsilon^{\mu\nu} \overline{\psi}_{+\mu} \psi_{-\nu} - \frac{1}{2} \sigma \epsilon^{\mu\nu} \overline{\psi}_{-\mu} \psi_{+\nu} \,.
  \end{aligned}
\end{equation}

\subsubsection{Chiral and anti-chiral multiplets}
\label{sec:sugra-chiral-anti}
\begin{equation}
  \begin{aligned}
\delta\phi&=   + \epsilon_+\psi_--\epsilon_-\psi_+\,,
\qquad\delta\overline\phi
= -\overline   \epsilon_+\overline \psi_- + \overline \epsilon_- \overline \psi_+\,,
\\
\delta\psi_-&=\epsilon_- F- \overline{\epsilon}_-\left({\rm z}-\sigma -\frac{q}{2}\widehat{\mathcal H}\right)\phi +2i\overline{\epsilon}_+ \mathcal{D}^\text{SG}_{\hat{z}}\phi\,,
\\
\delta\psi_+&=\epsilon_+ F + \overline{\epsilon}_+\left(\overline{\rm z}-\overline\sigma -\frac{q}{2}\widehat{\overline{\mathcal H}}\right)\phi +2i\overline{\epsilon}_- \mathcal{D}^\text{SG}_{\hat{\overline z}}\phi\,,
\\
\delta\overline\psi_-&=\overline \epsilon_- \overline F - \epsilon_-  \overline\phi  \left(\overline{\rm z} -\overline \sigma -\frac{q}{2}\widehat{\overline{\mathcal H}}\right) - 2i \epsilon_+  \mathcal{D}^\text{SG}_{\hat{ z}}\overline \phi\,,
\\
\delta\overline\psi_+&=\overline \epsilon_+ \overline F + \epsilon_+ \overline\phi \left({\rm z} - \sigma -\frac{q}{2}\widehat{\mathcal H}\right) - 2i \epsilon_-  \mathcal{D}^\text{SG}_{\hat{\overline z}}\overline \phi\,,
\\
\delta F&=-\left(\overline{\rm z}-\overline\sigma-\frac{q}{2}\widehat{\overline{\mathcal H}}\right) \overline\epsilon_+\psi_- - \left({\rm z}-\sigma-\frac{q}{2}\widehat{\mathcal H}\right)\overline{\epsilon}_- \psi_+ 
\\
&\qquad\qquad
+ 2i  \overline{\epsilon}_+  \mathcal{D}^\text{SG}_{\hat z}\psi_+ - 2i \overline{\epsilon}_-   \mathcal{D}^\text{SG}_{\hat{\overline     z}}\psi_- -i (\overline{\epsilon}_+\overline{\lambda}_--\overline{\epsilon}_- \overline{\lambda}_+)\phi
\\
&\qquad\qquad\qquad
-q\epsilon^{\mu\nu}  (\overline{\epsilon}_+ \mathcal{D}_\mu \overline{\psi}_{-\nu} +\overline{\epsilon}_- \mathcal{D}_\mu \overline{\psi}_{+\nu} )\phi
\,,
\\
\delta \overline{F} &= +\epsilon_+ \overline\psi_-  \left ({\rm z}-\sigma-\frac{q}{2}\widehat{\mathcal H}\right) + \epsilon_-\overline\psi_+\left (\overline{\rm z}- \overline\sigma - \frac{q}{2}\widehat{\overline{\mathcal H}}\right)
\\
&\qquad \qquad
 +2i  \epsilon_+  \mathcal{D}^\text{SG}_{\hat z}\overline\psi_+ - 2i  \epsilon_-  \mathcal{D}^\text{SG}_{\hat{\overline z}}\overline{\psi}_- -i\overline\phi (\epsilon_+\lambda_- - \epsilon_- \lambda_+)\,.
\\
&\qquad\qquad\qquad
+q \epsilon^{\mu\nu} (\epsilon_+ \mathcal{D}_\mu \psi_{-\nu} + \epsilon_-\mathcal{D}_\mu \psi_{+\nu}  )\overline{\phi} \,.
  \end{aligned}
\end{equation}

\subsubsection{Twisted chiral and anti-chiral multiplets}
\label{sec:sugra-twisted-chiral-anti}

\begin{equation}
  \begin{aligned}
& \qquad \qquad \qquad \qquad 
\delta v = \overline{\epsilon}_+ \chi_- -\epsilon_- \overline{\chi}_+\,, \\
&\delta\chi_- = 2i \epsilon_+ \mathcal{D}^{\rm SG}_{\hat{z}} v +\epsilon_- E\,,
\qquad
\delta \overline{\chi}_+ = 2i \overline{\epsilon}_- \mathcal{D}^{\rm SG}_{\hat{\overline{z}}}v + \overline{\epsilon}_+ E\,,\\
&
\qquad \qquad \qquad 
\delta E = 2i \epsilon_+ \mathcal{D}^{\rm SG}_{\hat z} \overline{\chi}_+ -2i \overline{\epsilon}_- \mathcal{D}^{\rm SG}_{\hat{\overline{z}}} \chi_-\,,
  \end{aligned}
\end{equation}

\begin{equation}
  \begin{aligned}
&\qquad  \qquad  \qquad  \qquad  \delta \overline{v}  = - \epsilon_+ \overline{\chi}_- + \overline{\epsilon}_- \chi_+ \,, \\
&  \delta \overline{\chi}_-  = - 2i \overline{\epsilon}_+ \mathcal{D}^{\rm SG}_{\hat{z}} \overline{v} + \overline{\epsilon}_- \overline{E} \,, 
\qquad  \delta \chi_+  = - 2i \epsilon_- \mathcal{D}^{\rm SG}_{\hat{\overline{z}}} \overline{v} + \epsilon_+ \overline{E} \,, \\
&  
\qquad \qquad \qquad 
\delta \overline{E}  = 2i \overline{\epsilon}_+ \mathcal{D}^{\rm SG}_{\hat{z}} \chi_+ - 2i \epsilon_- \mathcal{D}^{\rm SG}_{\hat{\overline{z}}} \overline{\chi}_- \,.
  \end{aligned}
  \end{equation}
Here $\mathcal{D}^{\rm SG}_a = e_a{}^\mu \mathcal{D}^{\rm SG}_\mu$
denotes the derivative that is covariant with respect to SUSY%
\footnote{%
By SUSY covariance we mean that the SUSY variation of $ \mathcal{D}^{\rm SG}_\mu$ acting on a matter field does not contain derivatives of $\epsilon$ or $\overline{\epsilon}$.
} as well
as the bosonic gauge symmetries.
It is defined as 
\begin{equation}
\mathcal{D}^{\rm SG}_\mu :=  \hat{D}_\mu + i \delta(\psi_{\mu}, \overline{\psi}_{\mu})\,,
\end{equation}
where $\hat{D}_\mu$ uses the modified spin connection $\omega_\mu+K_\mu$, and $\delta(\psi_{\mu}, \overline{\psi}_{\mu})$ denotes the SUSY transformation with the parameters $(\epsilon_-,\epsilon_+, \overline{\epsilon}_-, \overline{\epsilon}_+)$ replaced by $(\psi_{-\mu},  \psi_{+\mu},  \overline{\psi}_{-\mu}, \overline{\psi}_{+\mu})$.

\subsection{SUSY covariant curvatures}
\label{sec:susy-covar-curv}
The following quantities are covariant with respect to supersymmetry: their variations depend on the SUSY parameters algebraically and do not involve their derivatives.

The graviphoton field strengths $\widehat{\mathcal{H}}$ and $\widehat{\overline{\mathcal{H}}}$ receive corrections that involve gravitini as in (\ref{H-hat-H-bar-hat}).
Similarly the gauge field strength $v_{\mu\nu}=\partial_\mu v_\nu - \partial_\nu v_\mu + i [v_\mu,v_\nu]$ is SUSY-covariantized as in (\ref{gauge-field-srength-SUSY-covariant}).
The SUSY covariantization of the Ricci scalar is
\begin{equation}
  \begin{aligned}
\widehat{R} &=
  R + 2\epsilon^{\mu\nu}   \partial_\mu K_\nu   + i \epsilon^{\mu\nu}  \psi_{\mu} [  \widehat{\mathcal{H}}  (1+\gamma) +  \widehat{\overline{\mathcal{H}}} (1 - \gamma) ]\overline{\psi}_{ \nu}
\\
&\qquad \qquad \qquad
+ 2i \epsilon^{\mu\nu}  \psi_{\mu}   \gamma_\nu \mathcal{D}_\rho \overline{\psi}_\sigma  \epsilon^{\rho\sigma} +2i \epsilon^{\mu\nu}   \overline{\psi}_{ \mu} \gamma_\nu \mathcal{D}_\rho \psi_\sigma \epsilon^{\rho\sigma} \,.
  \end{aligned}
\end{equation}
We also have
\begin{equation}
  \begin{aligned}
  \widehat{F}^{\rm R}_{\mu\nu} & = F^{\rm R}_{\mu\nu} + \frac{i}{2} \psi_{ [\mu}  [  \widehat{\mathcal{H}}  (1+\gamma) - \widehat{\overline{\mathcal{H}}} (1 - \gamma) ]\overline{\psi}_{\nu]}    \\
& \qquad\qquad - i \psi_{ [\mu}  \gamma_{\nu]}\gamma \epsilon^{\rho\sigma}  \mathcal{D}_\rho \overline{\psi}_\sigma 
 + i \overline{\psi}_{[\mu} \gamma_{\nu]}\gamma \epsilon^{\rho\sigma} \mathcal{D}_\rho \psi_\sigma \,.
  \end{aligned}
 \end{equation}
The gravitino field strengths $\epsilon^{\rho\sigma}\mathcal{D}_{\rho} \psi_{\sigma}$ and $\epsilon^{\rho\sigma}\mathcal{D}_{\rho} \overline{\psi}_{\sigma}$ that appear in various places are also SUSY covariant.

\subsection{Relations between multiplets}
\label{sec:relat-betw-mult}

\subsubsection{Twisted chiral and twisted anti-chiral multiplets from the gravity multiplet}
\label{sec:twist-chir-twist}

The following fields constructed from the gravity multiplet forms a twisted chiral multiplet:
\begin{equation}
  \begin{aligned}
&v^{\rm gr} =  \widehat{\mathcal{H}} \,, \qquad 
\chi^{\rm gr}_- = + 2 \epsilon^{\mu\nu} \mathcal{D}_\mu { \psi{}_{- \nu}} \,,
\qquad
\overline{\chi}^{\rm gr}_+  =  - 2  \epsilon^{\mu\nu} \mathcal{D}_\mu { \overline{\psi}{}_{+ \nu}}\,,
\\
&
\qquad \qquad \qquad \quad
E^{\rm gr}  =  - \frac{1}{2}\widehat{R}  -\epsilon^{\mu\nu} \widehat{F}^{\rm R}_{\mu\nu}-  \widehat{\mathcal{H}} \widehat{\overline{\mathcal{H}}} \,.
  \end{aligned}
\end{equation}
Similarly,
\begin{equation}
  \begin{aligned}
&v^{\rm gr}  =\widehat{\overline{\mathcal{H}}} \,, 
\qquad
\overline{\chi}^{\rm gr}_- = + 2 \epsilon^{\mu\nu} \mathcal{D}_\mu \overline{\psi}_{-\nu} \,, 
\qquad
\chi^{\rm gr}_+ = - 2 \epsilon^{\mu\nu} \mathcal{D}_\mu \psi_{+\nu} \,, \\
&
\qquad \qquad \qquad\quad
\overline{E}^{\rm gr} = -\frac{1}{2}\widehat{R} + \epsilon^{\mu\nu} \widehat{F}^{\rm R}_{\mu\nu} - \widehat{\mathcal{H}} \widehat{\overline{\mathcal{H}}} \,.
  \end{aligned}
  \end{equation}
form a twisted anti-chiral multiplet.

\subsubsection{Gauge multiplet from the gravity multiplet}
\label{sec:gauge-multiplet-from}

We can also construct a gauge multiplet (R-symmetry gauge multiplet) from the gravity multiplet:
\begin{equation}\label{R-gauge-from-grav}
  \begin{aligned}
&\  v^{\rm R}_\mu =  A^{\rm R}_\mu\,,    \qquad  \sigma^{\rm R} = \frac{1}{2} \widehat{\mathcal{H}} \,, \qquad \overline{\sigma}^{\rm R} = \frac{1}{2}  \widehat{\overline{\mathcal{H}}} \,, \\
&\lambda^{\rm R}_- =  + i \epsilon^{\mu\nu} \mathcal{D}_\mu { \psi{}_{- \nu}}  \,, \quad \lambda^{\rm R}_+ =   -  i \epsilon^{\mu\nu} \mathcal{D}_\mu \psi_{+\nu} \,, 
\\
& \overline{\lambda}^{\rm R}_- = -i \epsilon^{\mu\nu} \mathcal{D}_\mu \overline{\psi}_{-\nu} \,, \quad \overline{\lambda}^{\rm R}_+ = + i \epsilon^{\mu\nu} \mathcal{D}_\mu \overline{\psi}_{+\nu}\\
&\qquad \qquad \qquad D^{\rm R}  = -\frac{1}{4} \widehat{R}  \,.
  \end{aligned}
\end{equation}

\subsubsection{Twisted chiral multiplet from the abelian gauge multiplet}
\label{sec:sugra-twisted-chiral-abelian-gauge}

An  abelian gauge multiplet gives rise to twisted chiral and twisted anti-chiral multiplets, indicated by superscript ``ga'':
\begin{equation}
  \begin{aligned}
& v^{\rm ga} = \sigma \,, \qquad  \chi^{\rm ga}_- = -i \lambda_- \,, \qquad \overline{\chi}^{\rm ga}_+ =  i \overline{\lambda}_+ \,, \\
& \qquad E^{\rm ga} =   D+2i \widehat{v}_{\hat z \hat{\overline{z}}} -\widehat{\overline{\mathcal H}}\sigma
\,.
  \end{aligned}
\end{equation}
\begin{equation}
  \begin{aligned}
&    \overline{v}^{\rm ga}  = \overline{\sigma} \,, \qquad  \overline{\chi}^{\rm ga}_- = i \overline{\lambda}_- \,, \qquad \chi^{\rm ga}_+ = - i \lambda_+ \,, \\
&\qquad \overline{E}^{\rm ga} = D - 2i \widehat{v}_{\hat z \hat{\overline{z}}} -\widehat{\mathcal H} \overline\sigma
\,.
  \end{aligned}
\end{equation}

\subsubsection{Twisted chiral multiplet for a holomorphic function}
\label{sec:twist-chir-mult}

Given a holomorphic function $\widetilde{W}(v^{(j)})$ of scalars $v^{(j)}$ in several twisted chiral multiplets, we get a new twisted chiral multiplet
\begin{equation}
  v^{\widetilde{W}} :=  \widetilde{W}\,,
\quad
  \chi^{\widetilde{W}}_- :=  \chi^{(j)}_- \partial_j \widetilde{W}\,,
\quad
  \overline{\chi}^{\widetilde{W}}_+ :=  \overline{\chi}^{(j)}_+ \partial_j \widetilde{W}\,,
\end{equation}
\begin{equation}
  E^{\widetilde{W}} :=  E^{(j)} \partial_j \widetilde{W} + \chi^{(i)}_- \overline{\chi}^{(j)}_+ \partial_i \partial_j \widetilde{W} \,.
\end{equation}

\subsection{Supergravity actions}
\label{sec:supergravity-actions}

Let $e$ denote the determinant of the vielbein $e^a{}_\mu$.  
The actions 
\begin{equation}
  \int d^2x \, e\left(
E + 2\psi_{+\hat{z}}\overline{\chi}_+ - 2\overline{\psi}_{-\hat{\overline{z}}} \chi_-  + \widehat{\overline{\mathcal{H}}} v 
\right)
\end{equation}
and
\begin{equation}
  \int d^2x \, e\left(
\overline{E} + 2\overline{\psi}_{+\hat{z}}\chi_+ - 2 \psi_{-\hat{\overline{z}}} \overline{\chi}_-  +\widehat{\mathcal{H}} \overline{v} 
\right) \,.
\end{equation}
are invariant under SUSY transformations up to total derivatives.

\section{Renormalizing the FI parameter in flat space}
\label{FI-flat}

Since renormalization is a UV phenomenon, one should be able to understand it in flat space.
In this appendix we study the one-loop renormalization of the FI parameter in flat space.

We first note that the auxiliary field $D$ couples to the chiral multiplet only through the $-\overline{\phi}D\phi$ term in (\ref{action-chiral}).
For simplicity, let us turn off the central as well as the R-charges, and assume that $\sigma$ and $\Lambda$ are constant.
For small $D$ and in flat space we have
\begin{equation}
  Z_\text{1-loop}^\text{PV} \simeq Z_\text{1-loop}^\text{PV} \Big|_{D=0}\Big (1 +  \int d^2x \langle \overline{\phi} \phi \rangle_\text{PV} D \Big) \,.
\end{equation}
Here $ \langle \overline{\phi} \phi \rangle_\text{PV}$ is the two-point function at the coincident points, regularized by the Pauli-Villars method:
\begin{equation}
     \langle \overline{\phi} \phi \rangle_\text{PV} = \int \frac{d^2 k}{(2\pi)^2} \sum_J \frac{\epsilon_J b_J }{k^2+|a_J \Lambda+b_J \sigma|^2} \,.
\end{equation}
The integral converges absolutely because $\sum_J \epsilon_J b_J=0$.
We then find
\begin{equation}
  \begin{aligned}
         \langle \overline{\phi} \phi \rangle_\text{PV} & = -\frac{1}{2\pi} \sum_J \epsilon_J b_J \log | a_J \Lambda + b_J \sigma| \\
         &= \frac{1}{2\pi}{\log|\Lambda/\sigma| }- \frac{C_1}{2\pi} + \mathcal{O}(\Lambda^{-1}) \,.
  \end{aligned}
\end{equation}
Thus the effective FI coupling (the coefficient of $D/2\pi$ in $-\log Z$) is
\begin{equation}
r_\text{eff}=  r_0 (\Lambda) - {\log |\Lambda/\sigma|}+ C_1 \,.
\end{equation}
The dependence of $r_0$ on $\Lambda$ must be such that this combination is finte as $\Lambda \rightarrow \infty$.
We fix the finite part by demanding that out effective FI coupling $r_\text{eff}$ coincides with that found in \cite{Witten:1993yc}: $r_\text{eff}=r(\mu)  + \log(|\sigma|/\mu)$.
This gives the relation (\ref{FI-renormalization}) between the renormalized and the bare FI parameters.

Though our regularization method differs from that of \cite{Witten:1993yc}, the difference is accounted for by the counterterms, yielding the equivalent renormalized theory.

\bibliography{refs}

\end{document}